\documentstyle[epsfig]{aipproc}

\begin{document}
\title{Asymmetries from Semi-inclusive Polarized Deep Inelastic Scattering}

\author{Eva-Maria Kabu{\ss}\thanks{supported by the BMBF}}
\address{Inst. f\"ur Kernphysik, University of Mainz\\ Becherweg 45,
55099 Mainz}
\smallskip
\address{for the SPIN MUON COLLABORATION\thanks{
Contribution to the proceedings of DIS97 (Chicago)}
}
\maketitle
\vspace{-0.7cm}

\begin{abstract}
The analysis of semi-inclusive data from deep inelastic muon nucleon scattering
is presented. The resulting charged hadron asymmetries  are used to determine 
polarized valence and sea quark distributions. In addition, a new approach to 
derive inclusive asymmetries is discussed.
\end{abstract}

To investigate the spin structure of the nucleon SMC studies deep inelastic 
scattering of polarized muons off polarized protons and deuterons.
The SMC-NA47 experiment used the M2 muon beam of the CERN SPS at incident
energies of 100 and 190 GeV. Data with polarized proton and deuteron targets
were taken between 1992 and 1996 (see also \cite{emk_magnon},\cite{emk_lesqen}).
Scattered muons and charged hadrons are detected in an open forward spectrometer
and electron-hadron separation is achieved with the help of an iron scintillator
calorimeter.

In the standard analysis the spin structure function $g_1$ is
determined from the measured inclusive asymmetry.
The produced hadrons can be used in two ways:
Firstly, by requiring any hadron in addition to the scattered muon a true
deep inelastic scattering is signaled and thus a discrimination between 
these events and background events like radiative events or elastic electron
muon scattering is achieved. Secondly, semi-inclusive asymmetries from 
positively and negatively charged hadrons can be used to determine the
polarized valence and sea quark distributions as a function of the momentum
fraction carried by the struck quark, $x$.

In the first approach a complementary method of determining inclusive 
asymmetries is being developed. In the standard analysis the measured photon
nucleon asymmetry is related to the total cross section by
$A_{\rm meas}^{\gamma \rm N}=\Delta \sigma_{\rm tot}/\sigma_{\rm tot}$ where 
 $$\sigma_{\rm tot}=\lambda \sigma_{1 \gamma}+\sigma_{\rm tail}^{\rm el}+
\sigma_{\rm tail}^{\rm coh}+\sigma_{\rm tail}^{\rm qel}
+\sigma_{\rm tail}^{\rm inel}$$
 $$\Delta \sigma_{\rm tot}=\lambda \Delta \sigma_{1 \gamma}+
\Delta \sigma_{\rm tail}^{\rm el}+
\Delta \sigma_{\rm tail}^{\rm coh}+
\Delta \sigma_{\rm tail}^{\rm qel}+\Delta \sigma_{\rm tail}^{\rm inel}$$
where $\sigma_{\rm tail}$ ($\Delta \sigma_{\rm tail}$) is the cross section from
radiative tails (elastic, coherent on nuclei, quasielastic and inelastic) and 
$\lambda$ is a polarisation independent multiplicative factor for the one-photon
cross section $\sigma_{1\gamma}$ ($\Delta \sigma_{1\gamma}$) accounting 
for vacuum polarisation and vertex corrections.

\begin{figure}
\begin{center}
\begin{tabular}[h]{ll}
\epsfig{file=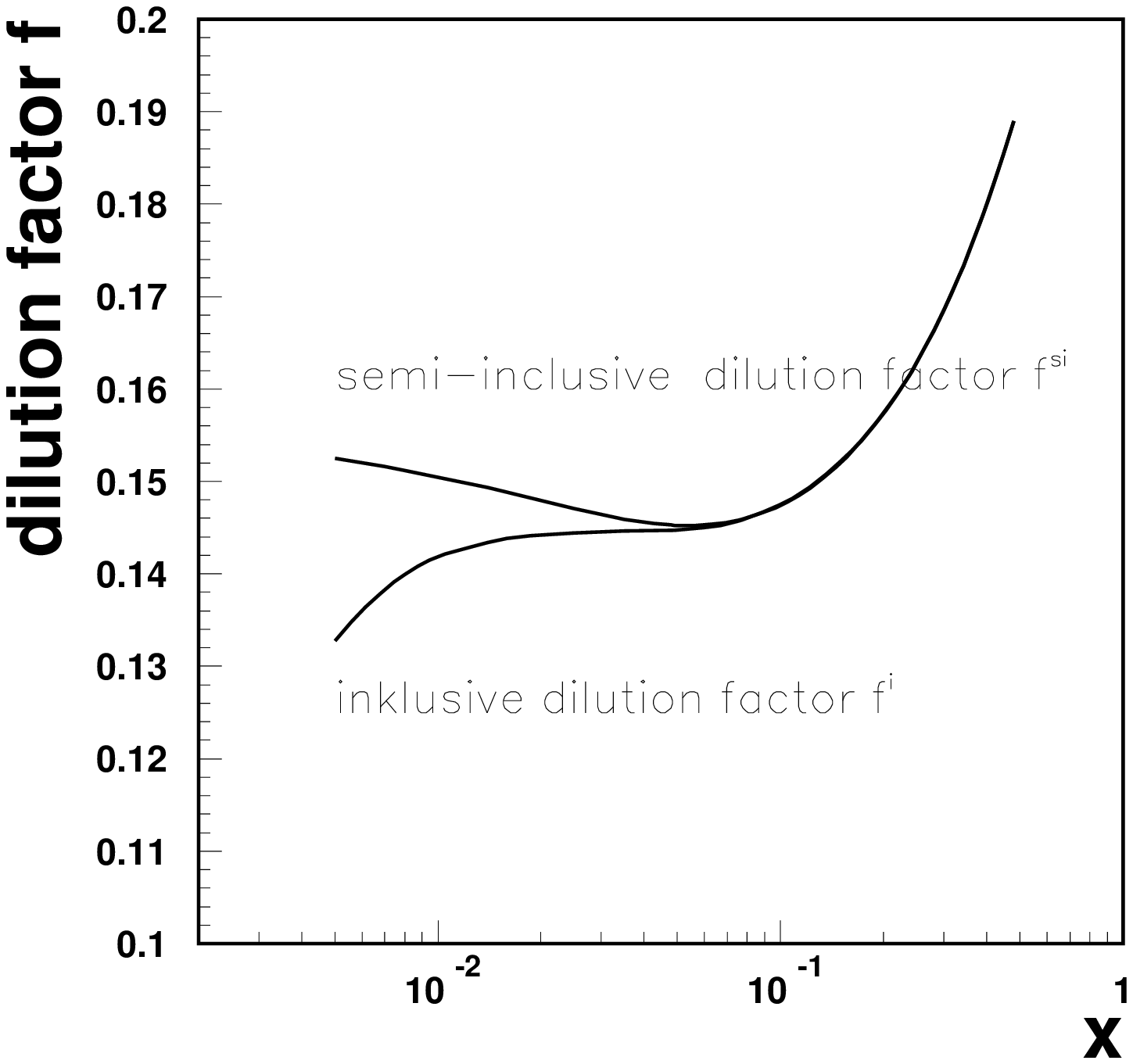,width=.4\textwidth} &
\epsfig{file=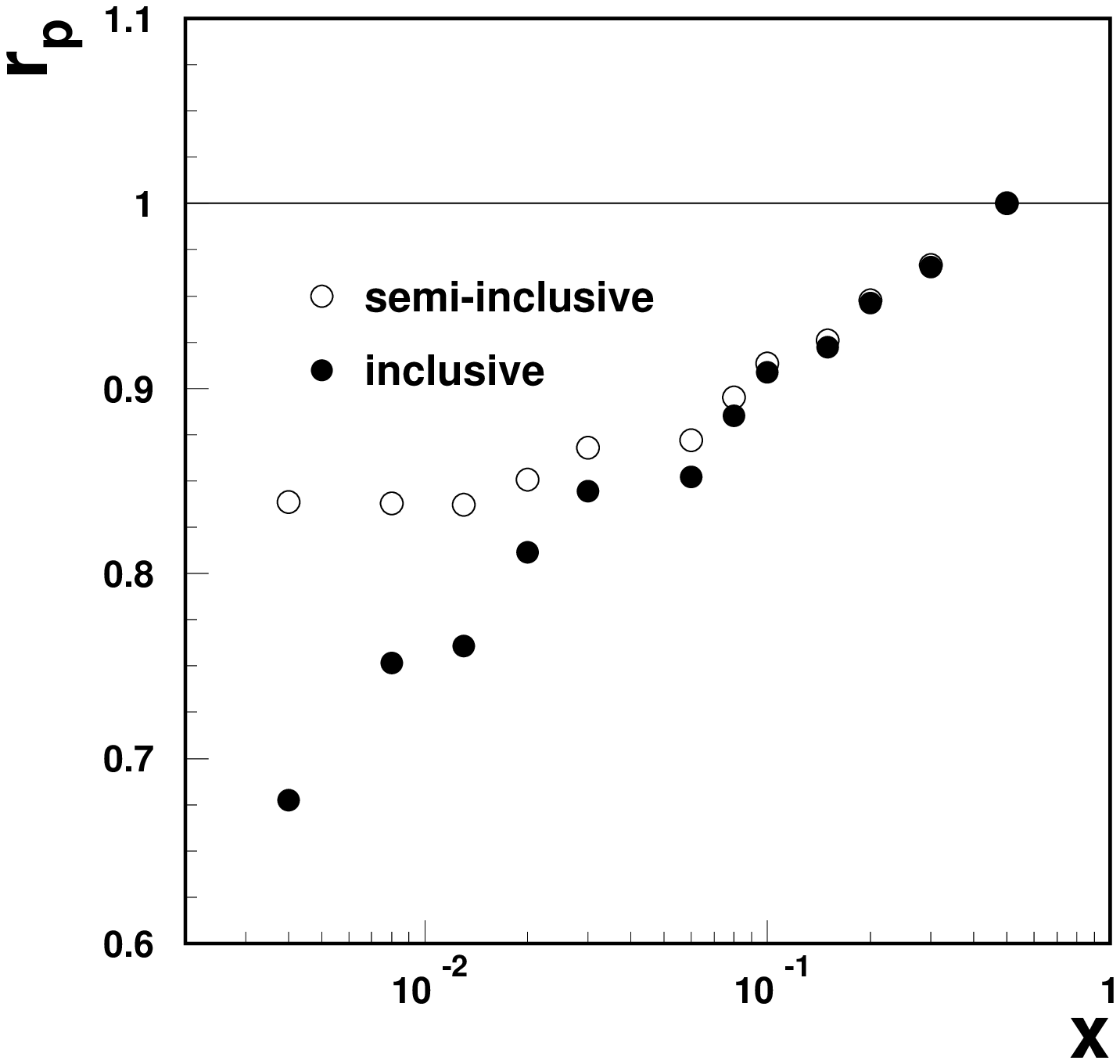,width=.4\textwidth}
\end{tabular}
\end{center}
\caption{Comparison of inclusive and semi-inclusive correction factor for
the dilution factor $f$ (left) for the ammonia target and the radiative 
correction factor $r_{\rm p}=\lambda \sigma_{1 \gamma}/\sigma_{\rm tot}$ for
the proton (right). }
\label{emk_fig_a1_corr} 
\end{figure}

Thus, the one-photon asymmetry 
$A_1^{\gamma \rm N}=\Delta \sigma_{1 \gamma}/\sigma_{1 \gamma}$ is diluted 
by these additional
processes and corrections have to be applied to determine $A_1^{\gamma \rm N}$
from 
$$A_1^{\gamma \rm N}=\frac{\sigma_{\rm tot}}{\lambda \sigma_{1\gamma}}
A_{\rm meas}^{\gamma \rm N} - \frac{1}{D} \cdot \frac{\Delta \sigma_{\rm tail}}
{\lambda \sigma_{1\gamma}},$$
where $D$ is the depolarisation factor for the virtual photon. An additional 
dilution is due to the presence of unpolarized nuclei in the target.
If an efficient selection of true deep inelastic events is obtained then
most of the contributions to $\sigma_{\rm tot}$ and $\Delta \sigma_{\rm tot}$ 
vanish
and the dilution of $A_1^{\gamma \rm N}$ is much smaller, giving a more accurate
measurement of $A_1^{\gamma \rm N}$ although the event sample is smaller. 
The change of the dilution factor $f$ and the radiative correction 
factor $r=\lambda\sigma_{1\gamma}/\sigma_{\rm tot}$ is illustrated in 
fig.\ref{emk_fig_a1_corr}.

In the SMC spectrometer mainly  forward produced charged hadrons with momenta
$p_{\rm h}>5$~GeV are detected, that is hadrons with
$z=E_h/\nu\approx 0.05$ with $\nu$ the energy transfer.
Events with a high mass $W$ of the final state contain in general several 
hadrons with high momenta. High $W$ events are typically found at small $x$,
so that a good acceptance is obtained for $x<0.1$, whereas in the high $x$ region
many hadrons are not detected due to the low momenta at low $W$. Monte Carlo
studies show that more than 80\% of the deep inelastic events are found
by requiring at least one additional hadron to the scattered muon.

For part of these hadrons that pass through the calorimeter also the ratio
$e=E_{\rm em}/E_{\rm tot}$ between the energy deposited in the electromagnetic
part to the total deposited energy in the calorimeter is measured. Particles
with $e>0.8$ are labelled electrons, with $e<0.8$ hadrons.

Electrons may either stem from $\gamma$ conversion after $\pi^0$ decays or
from radiative photons. Therefore, additional cuts were developed to remove
the latter ones keeping most of the electrons from $\pi^0$ decays.
Electrons with $z>0.2$ and $\alpha<0.004$ for events with $y=\nu/E_{\mu}>0.6$ 
were removed
with $\alpha$ the angle between the electron and the photon reconstructed
from the muon kinematics. The same cut is applied to particles that could not
be identified by the calorimeter due to the limited acceptance close to the
muon beam or due to multiple tracks pointing to the same module.
It was checked with Monte Carlo simulations that a possible bias for the
resulting asymmetries is much smaller than the statistical error and will
be included in the systematic error.

\begin{figure}
\begin{center}
\begin{tabular}[h]{ll}
\epsfig{file=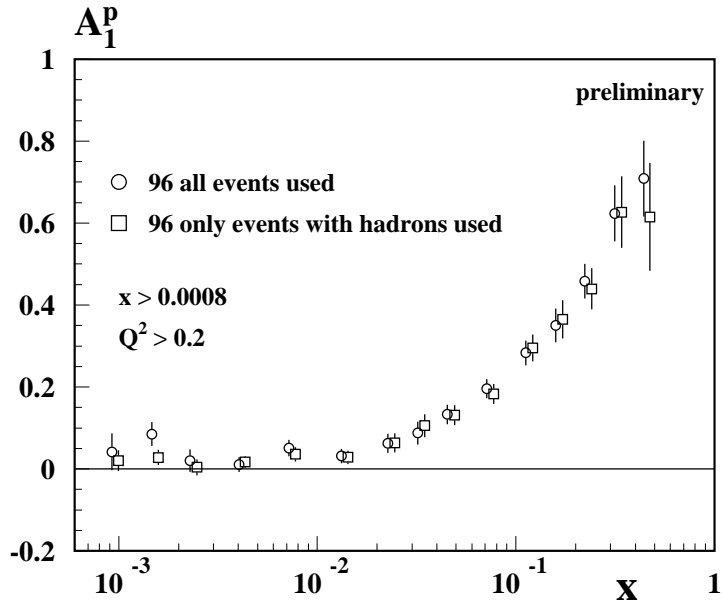,width=.63\textwidth}
\end{tabular}
\end{center}
\caption{ Preliminary results for $A_1^{\rm p}(x)$ obtained with the hadron method ( open
squares) compared to the ones from the standard analysis (open circles). The error
bars are statistical errors.}
\label{emk_fig_a1_had_meth} 
\end{figure}

The results for $A_1^{\rm p}$ using the 1996 data taken with an ammonia target
are shown  in fig.\ref{emk_fig_a1_had_meth} for $x>0.0008$ and $Q^2>0.2$~GeV$^2$
and are compared to the results from the standard analysis \cite{emk_magnon}.
Good agreement between the two results is obtained with the statistical error 
bars lower by a factor 0.6 to 0.9 for $x<0.01$ for the hadron method (for 
details, see \cite{emk_ewa}).

In the second approach polarized quark distributions are 
determined from semi-inclusive asymmetries using the different weighting of 
quarks
and antiquarks due to favoured and unfavoured fragmentation functions 
$D^{\rm h}_{\rm q (\bar{q})}$. In the quark parton model these hadron asymmetries
are given by
$$A^{\rm h}(x,z)=\frac{ \sum\limits_{\rm q} e_{\rm q}^2 (\Delta q(x) \cdot 
D_{\rm q}^{\rm h}(z) + \Delta \bar{q}(x) \cdot D_{\rm \bar{q}}^{\rm h}(z))}
{ \sum\limits_{\rm q} e_{\rm q}^2 (q(x) \cdot 
D_{\rm q}^{\rm h}(z) + \bar{q}(x) \cdot D_{\rm \bar{q}}^{\rm h}(z))}$$
with the unpolarized and polarized quark distributions, $q(x)$ and  $\Delta q(x)$.
Using parametrisations for the unpolarized distributions (GRV94 LO) \cite{emk_grv}
and the fragmentation functions from \cite{emk_emc} the polarized quark 
distributions can be extracted from a combined analysis of the inclusive
\cite{emk_magnon},\cite{emk_deut_g1} and the semi-inclusive asymmetries
\cite{emk_smc_hadrons}.

\begin{figure}
\begin{center}
\begin{tabular}[h]{ll}
\epsfig{file=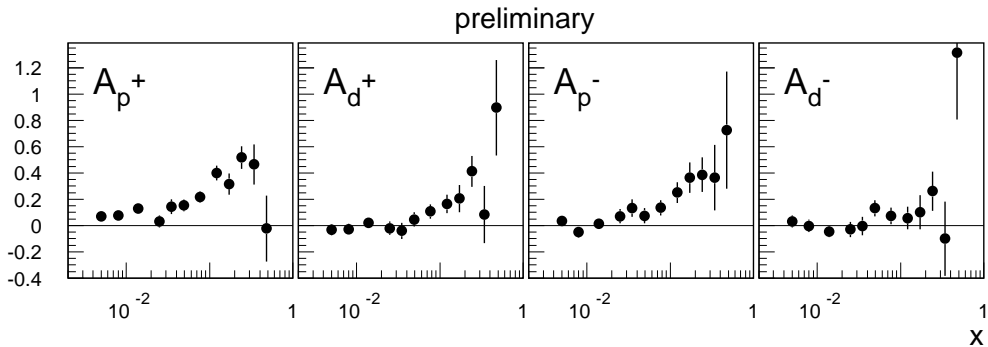,width=\textwidth}
\end{tabular}
\end{center}
\caption{Preliminary results for the semi-inclusive photon asymmetries for positively and
negatively charged hadrons from proton and deuteron targets as a function of 
$x$ with statistical error bars. }
\label{emk_fig_had_asy} 
\end{figure}

From all SMC measurements the asymmetries for positively and negatively 
charged hadrons were determined using tracks originating from
the interaction point of the scattered muon 
eliminating tracks that were labelled as electrons. To allow the interpretation
in the quark parton model a cut of $Q^2>1$~GeV$^2$ and $z>0.2$ was applied. 
The resulting
asymmetries are shown in fig.\ref{emk_fig_had_asy} in the $x$ range from 0.003 to 
0.7. The average $Q^2$ is 10 GeV$^2$ and 
the total event sample includes $5 \cdot 10^6$ positive and $3.8 \cdot 10^6$
negative hadrons.

To extract the valence quark distributions $\Delta u_{\rm V}(x)$ and
$\Delta d_{\rm V}(x)$ and the sea quark distribution $\Delta \bar{q}(x)$
additional assumptions are needed. Firstly, 
$\Delta \bar{u}(x)=\Delta \bar{d}(x)=\Delta \bar{q}(x)$ is used due to the limited
statistics of the data set. Secondly,
to constrain the polarized strange quark distribution  
$\Delta \bar{s}(x)=\Delta s(x)\sim s(x)$ is assumed with the integral 
$\int (s(x)+\bar{s}(x)) {\rm d}x=-0.1$ in the range $0.003<x<1$ fixed to the 
value from the inclusive analysis
\cite{emk_proton}. The effect of this assumption on $\Delta u_{\rm V}(x)$,
$\Delta d_{\rm V}(x)$ and $\Delta {\bar q}(x)$ is negligible due to the low
sensitivity of the measured asymmetries to the strange distribution as 80\%
of the hadrons are pions. Also, possible $Q^2$ dependences of the asymmetries
are neglected by using $A(x,Q^2_{\rm meas})=A(x,10\,{\rm GeV}^2)$.

\begin{figure}
\begin{center}
\begin{tabular}[h]{ll}
\epsfig{file=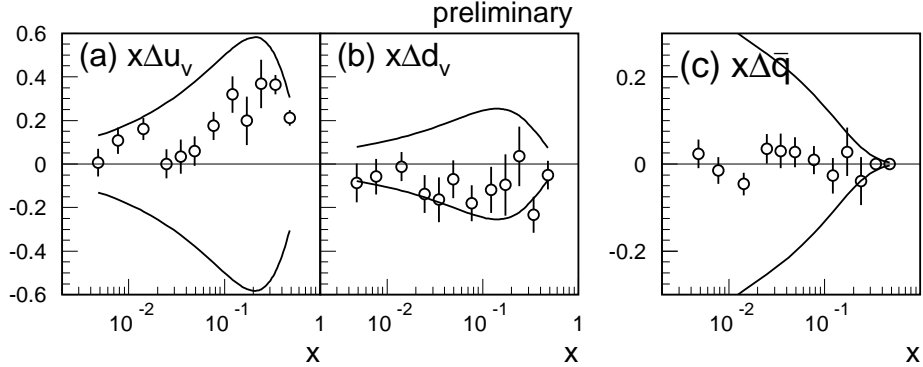,width=1.05\textwidth}
\end{tabular}
\end{center}
\caption{Preliminary results for the polarized quark distributions (a) 
$x\Delta u_{\rm V}$, (b) $x\Delta d_{\rm V}$ and (c) $x\Delta \bar{q}$
as a function of $x$ at $Q^2=10$~GeV$^2$ with statistical error bars. 
For the curves, see text.}
\label{emk_fig_dq} 
\end{figure}

The preliminary results for the polarized quark distributions $x\Delta q(x)$
are shown in
fig.\ref{emk_fig_dq} compared to $\pm xq(x)$ \cite{emk_grv}. The valence quark
distributions are positive for $\Delta u_{\rm V}(x)$ and negative for
$\Delta d_{\rm V}(x)$ with a polarization of about 50\% (-50\%), respectively.
The non-strange sea quark distribution is consistent with zero over the 
measured range.
For the integrals 
$\int\limits_{0.003}^{0.7} \Delta q {\rm d}x=\Delta q (0.003,0.7)$ in the measured range 
\begin{eqnarray}
  \Delta u_{\rm V}(0.003,0.7)=  0.88\pm 0.11
({\rm stat}.) \pm 0.06({\rm syst.}),   \nonumber \\
  \Delta d_{\rm V}(0.003,0.7)= -0.48\pm 0.15
({\rm stat}.) \pm 0.05({\rm syst.}),  \nonumber  \\
  \Delta \bar{q}(0.003,0.7) = -0.01\pm 0.05({\rm stat}.)
\pm 0.01({\rm syst.})  
\nonumber
\end{eqnarray}
are obtained. The extrapolation to $x=1$ leads to negligible contributions,
while the extrapolation towards low $x$ is currently under investigation.

\end{document}